\documentclass[12pt,preprint]{aastex}

\usepackage{natbib}

\newcommand{\pd}{\partial}

\newcommand{\mybf}{}

\usepackage{graphicx}

\begin{document}
\author{A. Dorodnitsyn\altaffilmark{1,2}, T. Kallman\altaffilmark{1}, G.S. Bisnovatyi-Kogan\altaffilmark{3}}
\altaffiltext{1}{Laboratory for High Energy Astrophysics, NASA Goddard Space Flight Center, Code 662, Greenbelt, MD, 20771, USA}
\altaffiltext{2}{Department of Astronomy/CRESST, University of Maryland, College Park, MD 20742, USA}
\altaffiltext{3}{Space Research Institute, 84/32, Profsoyuznaya st., Moscow, Russia}

\title{AGN obscuration through dusty infrared dominated flows. II. Multidimensional, 
radiation-hydrodynamics modeling}

\begin{abstract}

We explore a detailed model in which the active galactic nucleus (AGN) obscuration results from the extinction of AGN radiation in a global flow driven
by the pressure of infrared radiation on dust grains.
We assume that external illumination by UV and soft X-rays of the dusty gas located at approximately 1pc away from the supermassive black hole
is followed by a conversion of such radiation into IR. 
Using 2.5D, time-dependent radiation hydrodynamics simulations in a flux-limited diffusion approximation we 
find 
that the external illumination can support a geometrically thick obscuration via outflows driven by infrared radiation pressure in
AGN with luminosities greater than $0.05\, L_{\rm edd}$ and Compton optical depth,
$\tau_{\rm T}\gtrsim 1$.

  
\end{abstract}

\section{Introduction}
A fundamental assumption of active galactic nuclei (AGN) unification schemes is that type 1 and type 2 AGNs have similar intrinsic properties.
The basic premise of this paradigm is that obscuration and orientation effects are the major contributors to the observational dichotomy of AGNs.
The goal of this paper is to suggest an approach which explains the AGN dichotomy 
as resulting from the extinction of the AGN radiation in a hydrodynamical outflow powered by the pressure of the infrared radiation on the dusty plasma of AGN outskirts.

The suggestion that Seyfert 2 galaxies suffer from enhanced extinction compared to Seyfert 1 galaxies was made by
\cite{Rowan-Robinson77} based on the infrared observations.
However it was not until the seminal work of  \cite{Antonucci84}, and \cite{AntonucciMiller1985} when key evidence was collected from studies based on optical spectropolarimetry.  The detection of broad permitted lines in the polarized UV and optical spectrum of the nearby, luminous Seyfert 2 galaxy NGC 1068, confirmed that a bright, Seyfert 1 core is hidden behind optically thick, obscuring material.
Notice that a prediction of polarization of the X-ray flux in the $0.1-10$ keV range was made by \cite{Dorodnitsyn2010} based on theoretical
modeling of AGN outflows.

Direct evidence of the existence of toroidal obscuration comes from the mid- infrared observations of Seyfert 2 galaxies, such as 
the prototypical Seyfert 2 galaxy NGC 1068, and the closest AGN, the Circinus galaxy.
Observations of NGC 1068 using VLTI reveal a multi-component, multi-temperature dusty conglomerate: 
an inner, relatively small ($\sim 1$ pc) and hot ($\sim 800$ K) component embedded into an outer ($\sim 3.5$ pc) component which is much colder
($T\sim 320$ K) \citep{Jaffe2004, Raban09}.  In the Circinus galaxy
the observed elongated, 0.4 pc in diameter component is interpreted as a disk-like structure seen almost edge-on. 
This disk-like structure is co-incident with that inferred from the VLBI maps of $\rm H_{2}O$ maser emission \citep{Greenhill03},
being embedded into a much larger rounded component. This is interpreted as a geometrically thick torus with temperature $T\lesssim 300$K
\citep{Tristram07}.

The premise and the principal puzzle of AGN unification is the physical mechanism responsible 
for the geometrical thickness of the torus. 
Ample observational 
evidence for dust rules out support of the torus by gas pressure, as 
in such a case 
the temperature of the gas should be approximately  of the order of the virial temperature, $T_{\rm vir,g}=2.6 \times 10^{6}\,M_{7}/r_{\rm pc}$K, where $M_{7}$ is the black hole (BH) mass in $10^{7}\,M_{\rm \odot}$, and $r_{\rm pc}$ is the distance in parsecs.

Various mechanisms have been proposed to settle this issue: for example, in one of the first models a torus was considered being made of clumps having highly supersonic velocities \citep{KrolikBegelman88,BeckertDuschl04}. Magnetic fields are implicitly necessary in this model to provide enough elasticity to the clouds in order to avoid large dissipation though cloud-cloud collisions.
Another model suggested by \cite{Phinney89}, and \cite{Sanders89} considered a
locally geometrically thin, but globally warped disk.
Global magnetic fields were suggested to be a key ingredient either though hydromagnetic winds \citep{KoniglKartje94}, 
or as directly supporting the vertical balance of a quasi-static torus \citep{Lovelace98}. 
The main difficulty with hydromagnetic models comes from the large poloidal magnetic flux needed to support such a wind.
It is unclear whether such a strong global poloidal magnetic field exists at large distances from a BH.
{\mybf The clumpy nature of an outflow was addressed by \cite{ElitzurShlosman06}. These authors consider 
the dusty hydromagnetic obscuring wind as an alternative to quasi-static torus models.
}

It has been pointed out by \cite{PierKrolik92} that the infrared radiation pressure on dust may suffice to balance the vertical gravitational force.
Based on these ideas 
\cite{Krolik07}, \cite{ShiKrolik08} constructed a semi-analytic model of a static, infrared-supported torus. In all models in which infrared radiation is responsible for the torus thickness 
it is tacitly assumed that external radiation ranging from UV to soft X-rays is absorbed and converted into IR at the inner face of the torus. 

To sum up, previous models of AGN torus obscuration divide with respect to whether they are i) static, i.e. such as in a model of self-gravitating clouds or in a model
of a static IR supported torus; ii) dynamic, such as in hydromagnetic wind scenario. 

In this paper we present simulations which demonstrate that
obscuration at parses scale can be produced by global outflows driven by infrared radiation pressure on dust. 
Arguments for this model are both observational and theoretical.
Simple estimates show that the observed dust temperatures (see above) translate into high infrared radiation energy densities.  The latter when coupled with the high opacity of dust to IR radiation will produce strong radiation pressure force. Furthermore, one can easily see that radiation pressure on dust exceeds  the gas pressure and that together with gravity and centrifugal forces they determine the fate of the torus.
To support this line of arguments, in (\cite{Dorodnitsyn11a}, Paper I), it was shown that if the temperature in the torus exceeds that of  
$T_{\rm vir, r}   \simeq 312\,(n/10^{5} M_7 {r_{\rm pc}^{-1} })^{1/4} - 987 \,(n/10^{7} M_7 {r_{\rm pc}^{-1} })^{1/4}  \quad{\rm K}$, 
where $n$ is the number density, $M_{\rm BH}=10^{7}M_{\odot}$ is the black hole mass,
equilibrium between rotational, gravitational and radiation forces cannot be maintained resulting in the beginning of an outflow.
Paper I also presented a numerical solution of 2D transfer and dynamics subject to the restriction that the outflow is 1D and vertical.
{\mybf In Paper I it was assumed that external illumination by UV and soft X-rays and the subsequent conversion of this radiation into IR results in pumping of the torus with IR photons and producing a significant infrared pressure on dust.
A conservative scenario for the radiation acceleration was employed, that is only external sources were considered with no contribution from the accretion disk. It was also assumed that a thin accretion disk provides the necessary mass loading for the wind.
Such disk is buried in the dusty wind and its
characteristic temperature can be of the order of a few$\times 100-1000$K. Thus the disk contribution 
to the infrared pressure on dust can be significant. Given the highly speculative nature of estimates for the viscous transport in self-gravitating disks we 
do not take the disk contribution into account in our conservative estimate for the torus radiation driving in this paper. 

\cite{CzernyHryniewicz11} suggested the dusty wind as a possible origin for the low-ionization part of the broad line region. The effective temperature of the disk where the low ionization part of a broad line flow is formed in this picture is $T_{\rm eff} \sim1000 {\rm K} \gtrsim  T_{\rm vir, r}$. Thus, their broad line flow is the innermost part of the 
our obscuring flow. At these small radii the uplifted dusty flow is exposed to external heating, the dust evaporates, and the radiation force quenches which result in a failed wind scenario. The boundary between the two regions is roughly set by the dust sublimation radius $\sim 0.4 (L/10^{45})$pc, where $L ({\rm erg\, s^{-1}})$ is the total luminosity.
 }

The nature of the AGN obscuration problem calls for multidimensional simulations and reduce the predictive power of 1D modeling (which is extremely successful, for example, in stellar evolution calculations). 
Contrary to Paper I, in the current work we solve the full system of time-dependent equations of radiation hydrodynamics in three dimensions with axisymmetry (2.5D), and find the fate of the dusty
gas for conditions relevant to real AGNs.

The plan of this paper is as follows: 
In Section \ref{condtorus} we discuss some of the basic physical properties of dust and gas in the torus, and review
some of the results of Paper I which are extensively used in this paper. 
The torus problem can only be solved using radiation hydrodynamics (RHD). The basic setup of the equations of radiation hydrodynamics 
is explained in Section \ref{setup}. We also describe how physical conditions in a radiationally dominated plasma of a torus 
influence various regimes at which such system of equations can be solved.
In Section \ref{methods} we describe our numerical approach which we adopt in order to solve the full system of RHD. The implementation of the boundary conditions is also discussed.
A description of the results obtained from a calculated grid of models is given in Section \ref{results}.
We conclude with the discussion of the results and observational perspectives of our model in Section \ref{conclusion}.

\section{Radiation and matter in the torus}\label{condtorus}
At the inner parts of the accretion disk near the BH copious UV-,
and X-ray radiation is generated.
The very high opacity of dusty plasma to UV radiation makes impossible any static spherically-symmetric configuration if the UV luminosity, $L^{\rm UV}_{\rm c,dust}$
approaches the critical value (see Paper I):
	
\begin{equation}
L^{\rm UV}_{\rm c,dust}\simeq5\times 10^{-4}-0.01\, L_{\rm edd}\mbox{,}
\label{Lc_UVdust}
\end{equation}
where $L_{\rm edd}$ is the Eddington critical luminosity:

\begin{equation}
L_{\rm edd}=\frac{4 \pi c G M_{\rm BH}}{\kappa_{\rm T}}= 1.26 \times 10^{45}\,M_{7}\mbox{,}
\label{Lc_electrons}
\end{equation}
where $\kappa_{\rm T}=0.4\,{\rm cm^{2}\,g^{-1}}$ is the Thomson opacity due to electron scattering, and $M_{7}=M_{\rm BH}/(10^{7}M_{\odot})$.
Notice, that (\ref{Lc_UVdust}) is calculated taking $\kappa^{\rm UV}\simeq 6\times 10^{3} \kappa_{\rm T}$, the dust grain sizes of $0.025-0.25 \,\mu {\rm m}$ \citep{Mathis77},  grain density,  $n_{\rm d}= 2-3\, \rm g\,cm^{-3}$, and $50-100$ dust to gas mass ratio and assuming a perfect dust-gas coupling. 
For simplicity in the following $\kappa$ denotes the IR opacity of dust.
 
Analyzing the simplified model of the radiationally and rotationally supported torus, in Paper I it was found the following approximate condition 
for the temperature, $T$
for the beginning of an outflow:
\begin{equation}\label{outflow_cond2}
T_{0}>T_{\rm vir, r}(r_{0})\,\Gamma_{\rm c}^{1/4}\mbox{,}
\end{equation}
where $T_{0}$ is the temperature at the base of the wind at the distance $r_{0}$ from BH, and

\begin{equation}\label{Tvir_rad}
T_{\rm vir, r} = \left(\frac{GM_{\rm BH}\rho}{a r}\right)^{1/4} \mbox{,}
\end{equation}
where $\rho$ is the density,
$a=7.56\cdot 10^{-15}{\rm erg\, K^{-4} \, cm^{-3}}$ is the radiation density constant,
$\Gamma_{\rm c} = L/L_{\rm c}$, and $L_{c}$ is the critical luminosity in the infrared with respect to absorption on dust: 

\begin{equation}
L_{\rm c}=\frac{4 \pi c G M_{\rm BH}}{\kappa} \simeq (0.03 - 0.1)\,L_{\rm edd}\mbox{,}
\label{Le_ddington_dust}
\end{equation}
calculated assuming that the Rosseland mean opacity, $\kappa \simeq 10-30\kappa_{\rm T}$ in the temperature range $10^{2} -10^{3}$ K \citep{Semenov03}. 

The critical temperature  (\ref{Tvir_rad}) 
is, in fact, a definition of the virial temperature in a radiation-dominated plasma.
For densities relevant to numerical solutions presented in this paper $T_{\rm vir, r}$ spans from
 $312\,({M_{7}}/{r_{\rm pc}})^{1/4}$K for number density, $n=10^{5}$, to $987\,({M_{7}}/{r_{\rm pc}})^{1/4}$K for
 $n=10^{7}$.

In the dusty plasma of a torus, at typical values of $T$ and $\rho$ the pressure is dominated by the radiation. 
Its relative importance is described by the parameter $\displaystyle \beta=P_{\rm g}/P\simeq \left(10^{3} \,{ T^{3}_{3} }/ {n_{7} } +1\right)^{-1}$,
where $P$ is the total pressure 

\begin{equation}
\label{Pressure_total}
P=P_{\rm g}+ \Pi\mbox{,}
\end{equation}
where
\begin{equation}
P_{\rm g}=\frac{1}{\mu_{\rm m}}\rho{\cal R} T, \qquad \Pi=a\,T^{4}/3\mbox{,}
\label{Equations_of_State1}
\end{equation}
where $P_{\rm g}$, and  $\Pi$ are the gas and radiation pressures respectively,
${\cal R}=8.31\cdot 10^7 {\rm erg\, K^{-1} \, g^{-1}}$ is the universal gas constant, and $\mu_{\rm m}$ is the mean molecular weight. 
Thus, at typical densities $n=10^{6}-10^{7} {\rm cm}^{-3}$ and temperatures $T=10^{2}-10^{3}\rm K$ one finds
$P_{\rm g}\ll \Pi$.

When the approximate condition (\ref{outflow_cond2}) is met, a radiatively-driven wind develops. The terminal velocity, $v^{\infty}$ then can be estimated
assuming spherical symmetry and neglecting $P_{\rm g}$ in favor of the radiation pressure, $F_{\rm IR}\kappa/c$, obtaining 

\begin{equation}
\displaystyle v^{\infty}=\sqrt{ 2 GM/r_{0} \,  (\Gamma_{c} -1)  } \simeq 293 \, (M_{7} \Gamma_{0.5} \kappa_{10})^{0.5} \,
{\rm km\,s^{-1}  } \mbox{.}\label{vterm}
\end{equation}

The characteristic temperature of the conversion layer, $T_{\rm eff}$ was found in Paper I:

\begin{equation}\label{Teff00}
T_{\rm eff} =\left(4\alpha\Gamma\frac{GM_{\rm BH}}{\kappa_{\rm T} {a r^{2}}}\right)^{1/4} \simeq 463\,(\frac{\Gamma \gamma_{0.5}M_{7}} { r^{2}_{\rm pc}} )^{1/4}
\quad {\rm K} \mbox{,}
\end{equation}
where $\alpha\simeq 0.5$ is the fraction of the incident X-ray flux re-emitted in the IR into the torus, and 
$\Gamma = L/L_{\rm edd}$
In the calculations presented in this paper we 
adopt the parameter $\Gamma$, instead of $T_{\rm eff}$, 
calculating the latter from (\ref{Teff00}).

\section{The model setup and basic equations}\label{setup}
In the frame of reference co-moving with the fluid, the description of the interaction between radiation and matter is simplest, 
free from aberration and Doppler effects. For example, only in such a frame the adopted emissivity and absorptivity of matter are isotropic 
and only in such a frame they
correspond to those tabulated from laboratory experiments.
As such, it is important to distinguish between reference frames when casting equations for matter and radiation.
The equations of radiation hydrodynamics, to first order in $v/c$, can be formulated in the following form
\citep{MihalasBookRadHydro}:
\begin{eqnarray}
D_{t}\rho &+& \rho\,{\bf \nabla\cdot v}=0\mbox{,}\label{eq11}\\
D_{t}\bf {v} &=& -\frac{1}{\rho}\,\nabla p + {\bf g_{\rm rad}} -\nabla \Phi \mbox{,}\label{eq12}\\
\rho D_{t}\left(\frac{e}{\rho}\right) & = & - p{\bf \nabla\cdot v} - 4\pi\chi_{\rm P}B + c\chi_{\rm E}E\mbox{,}\label{eq13}\\
\rho D_{t}\left(\frac{E}{\rho}\right) & = & -{\bf \nabla\cdot F} - {\bf \nabla v : P} + 4\pi\chi_{\rm P}B- c\chi_{\rm E}E\mbox{,}\label{eq14}
\end{eqnarray}
where quantities related to matter: $\rho$, $p$, and $e$ are
the material mass density, gas pressure and gas energy density, ${\bf v}$ is the velocity;
quantities related to radiation are the frequency-integrated moments:  $E$ is the radiation energy density, 
$\bf F$ is the radiation flux, and $\bf P$ is the radiation pressure tensor; $\chi_{\rm P}$, $\chi_{\rm E}$ are the Planck
mean and energy mean absorption opacities (in $\rm cm^{-1}$),
; $c$ is the speed of light, $B=\sigma T^{4}/\pi$ is the Planck function, and 
$\sigma = a c/4$ is the Stefan-Boltzmann constant, $T$ is the gas temperature;
other notation include the convective derivative,
$ D_{t} = \frac{\partial}{\partial t} + \bf{v\cdot \nabla}$, and ${\bf \nabla v : P}$ in (\ref{eq13}) denotes
the contraction $(\partial_{j} v_{i}) P^{ij}$. Notice that all the dependent variables in (\ref{eq11})-(\ref{eq14}) are evaluated in the co-moving frame. 

Frequency-independent moments 
$E$, $\bf F$,  which appear in the above set of RHD equations are obtained by
calculating angular moments from the frequency-integrated specific intensity, $I({\bf r}, \Omega, \nu, t)$:

\begin{eqnarray}\label{moments1}
E(\bf{r},t) &=& \frac{1}{c}\int_{0}^{\infty}d\nu  \oint d\Omega \, I({\bf r}, \Omega, \nu, t)\mbox{,}\\
{\bf F}(\bf{r},t) &=& \int_{0}^{\infty}d\nu\oint d\Omega \,\hat{\bf n} \,I({\bf r}, \Omega, \nu, t)\mbox{.}
\end{eqnarray}
The frequency-independent radiation pressure tensor
$\bf P$ is found from

\begin{equation}\label{moments2}
{\bf P}({\bf r},t) = \frac{1}{c}\int_{0}^{\infty}d\nu\oint d\Omega \,\hat{\bf n} \hat{\bf n} \,I({\bf r}, \Omega, \nu, t)\mbox{.}
\end{equation}
The radiation force, ${\bf g_{\rm rad}}$ is calculated from the following relation	

\begin{equation}
{\bf g_{\rm rad}}	= \frac{1}{c}\frac{\bf \chi_{\rm F} \,{\bf F} }{\rho}\mbox{,}
\end{equation}
where $\chi_{\rm F}= \chi_{\rm a}+\chi_{\rm T}$ is the total flux mean opacity consisting of absorption opacity, $\chi_{\rm a}$ and the Thomson scattering opacity, $\chi_{\rm T}$. In the following we will not differentiate between $\chi_{\rm F}$, $\chi_{\rm P}$ and $\chi_{\rm E}$, and
omit subscript from $\chi$ where appropriate.

In addition to (\ref{eq11})-(\ref{eq14}), the full system of equations of radiation hydrodynamics should include an equation for ${\bf F}$ . 
However, we adopt a flux-limited diffusion approximation (FLD), and
there is no need of such equation as the closure relation between ${\bf F}$ and $E$ is found from the diffusion law:

\begin{equation}\label{eq1}
{\bf F}=-\frac{c}{3\kappa\rho}\,\nabla E = - D \, \nabla E \mbox{,}
\end{equation}
where 

\begin{equation}
D= c\, \lambda\mbox{,}\label{DifCoef}
\end{equation} 
is the diffusion coefficient and $\lambda$ is 
the photon mean free path: $\lambda=1/(\kappa \rho)$, where $\kappa = \chi/\rho$.

The diffusion approximation is adopted by tacitly assuming that optical depth $\tau \gtrsim 1$.
To take into account the possibility of $\tau<1$ the diffusion approximation should be modified so it has a correct limiting behavior .
Notice that in a free-streaming limit $|{\bf F}|\to c E$.
However, when $\tau \ll 1$ the mean free path, $\lambda \to \infty$,  and
$D\to \infty$, and $|{\bf F}|\to \infty$. That is, when optical depth becomes small, or when $\rho\to 0$, the standard diffusion approximation is no longer applicable.

In order to overcome this problem the standard approach is to adopt
the flux-limited diffusion approximation \citep{AlmeWilson74,Minerbo78,LevermorePomraning81}.
In an FLD approximation $\lambda$ is replaced by $\lambda^{*} = \lambda\,\Lambda$,
where $\Lambda$ is the flux limiter. The flux limiter we adopt in the current work is that of \cite{LevermorePomraning81}:

\begin{equation}\label{LP_FluxLim}
\Lambda = \frac{2+R_{\rm LP}}{6+3R_{\rm LP}+R_{\rm LP}^{2}}\mbox{,}
\end{equation}
where $R_{\rm LP}=\lambda\,|\nabla E |/E$. If $\tau\to 0$, then $R_{\rm LP}\to \infty$, and 
$|F| \sim c\,E$. In the optically thick limit $R_{\rm LP}\to 0$ and $\Lambda\to 1/3$.

The gas is assumed to be in a local thermodynamic equilibrium (LTE) at a temperature, which can be different from that of the effective temperature of 
radiation. The opacities are treated in a grey approximation, i.e. frequency-independent, constant dust opacity is assumed.
The equation of state is taken to be that of an ideal polytropic gas, $p=(\gamma-1)e$ with the ratio of
specific heats $\gamma$. The gas temperature is obtained from the relation $T=(\gamma-1)\mu_{\rm m}e/(\rho \cal R)$.
Given the great variety of physical conditions in dusty molecular gas, exposed to UV, and X-ray radiation, we set  $\mu_{\rm m}=1$ throughout this paper. 

Notice that in Paper I, radiation properties were calculated assuming stationarity and radiative equilibrium.
To understand why $\nabla\cdot {\bf F}=0$ may be not too bad an approximation
it is instructive to compare time scales of the variation of the radiation field.
In most cases we expect the radiation field to follow matter with the fluid-flow time scale, 
$t_{f}=l/v \sim l/v_{\rm k}\simeq 5\cdot 10^{3} \,\l_{1}^{3}\,M_{7}^{-1/2}\,  {\rm yr}$,
where $v_{\rm k}$ is the orbital velocity, $l$ is a typical length scale of the system, and $l_{1}=l/1{\rm pc}$.   
Another important scale is
the time which is necessary for a photon to travel distance $l$:
$t_{\rm r} = l/c \simeq 3.43 \,\l_{1}\, {\rm yr}$, for an optically thin case. In an optically thick case a photon travels across $l$ by means of a random walk.
The corresponding time scale in a diffusion regime is the diffusion time 
$t_{\rm d}=l^{2}/c\lambda = t_{\rm r}\, l/\lambda\simeq 530\,\l^{2}_{1} n_{6}\kappa_{10}\, {\rm yr}$ 

If $t_{\rm r} \ll t_{\rm f}$ the radiation field adjusts almost instantaneously to changes of the physical conditions in the flow.
Consequently, calculations are simplified immensely because the explicit time
variation of the radiation field can be ignored.
At a given time step the properties of the flow can be calculated taking into account the radiation as it is frozen i.e. 
at the flow time scale the radiation is essentially a sequence of snapshots which instantly adapts to the flow.

If additionally, optical depth $\tau$ is sufficiently high then the radiation-matter energy exchange term in (\ref{eq13}), and  (\ref{eq14}) scales as
$B- cE/4\pi \simeq {\cal  O} (B/\tau^{2})$. If this condition is augmented by a requirement of a strict stationarity: $\partial / \partial t =0$,
the resultant situation is described as a radiative equilibrium and 
the radiation field can be found from the equation $\nabla \cdot {\bf F}=0.$ As was mentioned, the latter was used in Paper I to find radiation field in a stationary outflowing torus.  

Contrary to Paper I, in the present work we do not {\it assume} the existence of the wind nor are we concerned with a wind solution which is stationary in a strict sense (both are the assumptions of Paper I). 
Rather, we solve for the full time dependence of both the radiation field and the flow.
The effective coupling outlined above between time scales and local optical depth (see e.g. \cite{BKBlinn78}) presents a significant challenge to any RHD simulations.
Careful analysis of the RHD system of equations \citep{MihalasBookRadHydro} prescribes that
in order to be consistent in all regimes of the radiation-matter interaction, including the ability to describe correctly regions of the flow where $\tau < 1$ or
$\tau \gg 1$, all terms in the system of RHD (\ref{eq11})-(\ref{eq14}) must be retained.

\section{Solution: method}\label{methods}

\subsection{Nondimensionalization} 
In order to explicitly extract governing parameters it is convenient 
to convert (\ref{eq11})-(\ref{eq14}) into a dimensionless form adopting dimensionless variables:
${\tilde r}=r/R_{0}$, ${\tilde t}=t/t_{0}$, where $R_{0}$, is a fiducial distance from the BH, $t_{0}=R_{0}/v_{0}$ is the characteristic flow-time, and ${\tilde v}=v/v_{0}$, where $v_{0}=\phi_{0}^{1/2}$,
and $\phi_{0}=GM_{\rm BH}/R_{0}$;
matter variables convert as ${\tilde \rho}=\rho/\rho_{0}$, ${\tilde p}=p/e_{0}$, ${\tilde e}=e/e_{0}$, where $\rho_{0}=n_{0}m_{\rm p}$ is the fiducial mass density and
$n_{0}$ is the number density, $m_{\rm p}$ is the proton mass, and $e_{0}= \rho_{0} v_{0}^{2}$; 
radiation variables transform as
${\tilde E}=E/E_{0}$, ${\tilde {\bf F} }={\bf F}/cE_{0}$, and ${\tilde {\bf P} }={\bf P}/E_{0}$, where $E_{0}=aT_{0}^{4}$, where $T_{0}$ is the fiducial temperature. The opacity transforms as ${\tilde \chi}=\chi/\chi_{0}$, where $\chi_{0}=1/R_{0}$.
Using such nondimensionalization and further simplifying notation by (hereafter) omitting tilde, we obtain

\begin{eqnarray}
D_{t}\rho &=& -\rho\,{\bf \nabla\cdot v}\mbox{,}\label{eq21}\\
\rho \, D_{t}\bf{v}  &=& -\nabla p +   A_{1} \, \chi \,{\bf F} -\rho\nabla \Phi  \mbox{,}\label{eq22}\\
\rho D_{t}\left(\frac{E}{\rho}\right) & = & -\frac{1}{\beta_{0}} {\bf \nabla\cdot F} - {\bf \nabla v : P} + 
\frac{1}{\beta_{0}} \kappa\rho \,(T^{4} -E) \mbox{,}\label{eq23}\\
\rho D_{t}\left(\frac{e}{\rho}\right) & = & - p{\bf \nabla\cdot v} - \frac{1}{\beta_{0}} A_{1}\kappa\rho \,(T^{4} -E)\mbox{,}\label{eq24}
\end{eqnarray}
where $\beta_{0} = v_{0}/c$, and non-dimensional  parameters are

\begin{equation}\label{param1}
A_{1}= \frac{E_{0}}{e_{0}} = \frac{aT_{0}^{4}}{\rho_{0} v_{0}^{2}} \mbox{,} \quad A_{2}=(\gamma-1)\frac{e_{0}\mu_{\rm m} } { \rho_{0} {\cal R} T_{0}}
\mbox{.}
\end{equation}

We solve equations (\ref{eq21})-(\ref{eq24}) adopting a time-dependent, axisymmetric $2.5D$ approximation,
meaning that we keep track of all $\phi$- components of all vector quantities, such as rotational velocity, $v_{\phi}$, but assume $\partial/\partial\phi\equiv 0$.
A cylindrical ${z,R}$ coordinate system is adopted in all our computations.

We derive our code from the family of the ZEUS codes \citep{Stone92a}. The hydrodynamical part of our code partially adopts methods and infrastructure found in ZEUS-2D, and
ZEUS-MP codes \citep{Hayes06}, including our original modifications summarized in 
\cite{Dorodnitsyn08b}.
Our original implementation of the radiation module which is adopted in the current work is built upon a simplified version developed in Paper I.

Notice that the hydrodynamical part of the ZEUS codes is based on a two-step update of the dependent variables: In the source step the local update is done for $\partial_{t} {\bf v}$, 
to account for various forces: gas pressure, $(\nabla p)/\rho$, radiation pressure ${\bf g}_{\rm rad}$, and the gravitational force $\nabla\phi$;
$\partial_{t} {e}$ is updated taking into account $p\cdot\nabla {\bf v}$ tern, i.e. $p\,dV$ work, where $V$ is the specific volume; shocks are treated adopting the artificial viscosity prescription of
\cite{NeumannRichtmyer50}. Corresponding viscous stresses and dissipation due to artificial viscosity, are also added in the source step.
Next is the transport step: the previously updated quantities are further updated taking into account fluid advection.

According to the general strategy adopted in the ZEUS  codes the update of the radiation energy is also made adopting the operator splitting of equation
(\ref{eq23}) into source and transport terms. 
The most time consuming part  is the solution of the equation (\ref{eq23}) for $E$ in the source step:
Thus, the radiation source step includes the finite difference update of $E$ from

\begin{eqnarray}\label{divE1}
\frac{\pd E}{\pd t} &=& - \beta_{0}^{-1} {\bf \nabla\cdot F} = \beta_{0}^{-1} \nabla\cdot (D\nabla E)\nonumber\\
&=&\beta_{0}^{-1}\left( \frac{\partial}{\partial l_{1}}(D (\nabla E)_{z} ) + \frac{\partial}{\partial l_{2}}(x D (\nabla E)_{x} ) \right)\mbox{,}
\end{eqnarray}
Equation (\ref{divE1}) is solved numerically adopting an alternative direction implicit scheme (ADI) \citep{Fletcher_book,
Fedorenko_book}. 
{\mybf The fact that the effective time step in (\ref{divE1}) is $\sim c/v$ times larger than the original time step demands the sub-cycling
when advancing (\ref{divE1}) over the original time-step $dt$. 
This is done by splitting $dt$, into 10-1000 sub-steps.}

Since in most (but not all!) of the dense parts of the flow $t_{\rm r}, t_{\rm d} \ll t_{\rm f}$ one obtains that
solving for the ${{\bf \nabla\cdot F} \simeq 0}$ term is the most important part of the solution for the radiation field at a given time step.
A three-point finite differencing stencil is used to approximate the diffusion operator in (\ref{divE1}), which involves a solution of a tri-diagonal matrix equation. 
Further details of how the $\nabla \cdot {\bf F}$ term in equation (\ref{divE1}) is taken into account can be found in Paper I. 

Our radiation module adopts and enhances the radiation module described in Paper I. 
The latter was designed to integrate equation $0= \partial E/ \partial w = -{{\bf \nabla\cdot F} \simeq 0}$ over a pseudo-time $w$ towards a stationary solution. This extensively tested module was augmented by the algorithms needed to take into account ${\bf \nabla v : P} + 
\frac{1}{\beta_{0}} \kappa\rho \,(T^{4} -E)$ terms in equation (\ref{divE1}). 
This part of the solution implements the well tested algorithms found in \cite{TurnerStone01} version of the ZEUS code.

The numerical grid used in the radiation module adopts the same staggered grid used in the hydrodynamic part of the ZEUS code. On such a grid, for example, $E$ is placed cell-centered,
while $D$ is face centered. 
Finite differencing in (\ref{divE1}) is done by approximating the derivatives making use of volume elements $dl_{2} = x\,dx$ and $dl_{1}=dz$. This approach allows to avoid approximation errors near the coordinate singularities \citep{StoneNorman92}.

An important part of the method is the implementation of the implicit time-stepping algorithm, and 
from this, the most important is that in our method during the update from $E^{n}_{ij}$ to $E^{n+1}_{ij}$
the diffusion coefficients $D_{ij}$
are taken at the ''old time'',  $t^{n}$. The method we are using for the radiation-matter interaction part adopts that of \cite{TurnerStone01}.
This includes a simultaneous solution of the following finite difference algebraic system of equations obtained from (\ref{eq23})-(\ref{eq24}):

\begin{eqnarray}
E^{n+1}-E^{n} &=&\left(-({\bf \nabla v : P})^{n+1} + \beta_{1}\,\kappa^{n}\rho^{n} ((T^{n+1})^{4} - E^{n+1})  \right)\delta t\mbox{,}\label{radgas-update1}\\
e^{n+1}-e^{n} &=&\left(-(p{\bf \nabla \cdot v })^{n} - A_{1} \beta_{1}\,\kappa^{n}\rho^{n} ((T^{n+1})^{4} - E^{n+1}) \right)\delta t \mbox{,}\label{radgas-update2}\\
\end{eqnarray}
where  $\delta t=t^{n+1}-t^{n}$ is the time step, and the omitted subscript ${ij}$ is assumed for all dependent variables, and other notation is introduced: $\beta_{1} = \beta_{0}^{-1}$, 
$T^{n+1}\equiv (T^{n+1})  = A_{2} (e^{n+1})/\rho^{n}$.

Benefits from such a discretization of variables in time are twofold: 
i) no linearization of difference equations is  required, as opposed to a fully implicit method;
ii) it allows 
for a much simpler update of the 
radiation and gas energies at the radiation-matter interaction step, in which case equations (\ref{radgas-update1})-(\ref{radgas-update2})
are reduced to a single scalar algebraic equation \citep{TurnerStone01}.
The numerical solution of the latter is much more robust compared to what is needed in a fully implicit method  (e.g. such as in \cite{Hayes06}).

{\mybf In
our hydrodynamic framework we combine the original radiation diffusion and boundary conditions module (aka the ''radiation module''), the original matter-
and radiation-energy update module and the hydrodynamical module. The hydrodynamical part is taken from that of ZEUS-MP framework without alteration.
This includes radiation advection step (transport step in zeus terminology) and the time-step calculation machinery. This part of the ZEUS-MP
has been extensively tested by \cite{Hayes06} against the Marshak wave test and the radiation-dominated shock wave test.

The radiation module developed in Paper I was verified against the time evolution of the solution with only the diffusion operator in (\ref{eq23}), the interaction between radiation and matter via matter-coupling terms in (\ref{eq23}),(\ref{eq24}), and the verification of the interface between the radiation terms in the rest of the zeus hydrodynamical algorithms. 

The tests we have implemented are those described in detail in \cite{TurnerStone01}, and we outline them in the following. In the first test we follow how the static uniform matter initially out of balance with radiation approaches 
thermal equilibrium. Assuming that $E$ is constant (notice that in the torus $E\gg e$), equation (\ref{eq24}) is reduced to an ordinary differential equation (ODE) for unknown $e$. Solving this ODE we compare the resultant solution with the full solution from the radiation module and thus verify the radiation-matter energy update in the code. We found an excellent agreement $(| e_{\rm code} - e_{\rm a}|/E_{\rm a} <1\%)$, where $e_{\rm a}$ is the solution of the ODE.

The second major test involves the evolution of the radiation flux divergence term alone. Generally, the easy way to do it is to compare the numerical solution in the optically thick regime with the analytical solution 
of the heat diffusion equation. 
The idea is that one can write an analytic solution (such as \citep{TurnerStone01}, eq. (47)) of the diffusion equation with constant diffusion coefficients on the unit square with periodic boundary conditions, and to compare it with the solution from the diffusion solver (with no hydrodynamics). 
This was done in Paper I where a very good agreement was obtained between such an analytic solution and the numerical solution of the 2D diffusion equation.

}

\subsection{Boundary conditions and the grid}

The boundary conditions (BC) are adopted from Paper I where one can find a corresponding discussion.
{\mybf 
Cylindrical $\{z,R\}$ coordinates were adopted.
}
The computational domain spans from $R_{0}$ to $R_{1}$ in horizontal, and from $z_{0}=0$ to $z_{1}$ in vertical direction in the 
meridional plane. 
{\mybf 
The stiffness imposed by radiation terms together with the necessity to sub-cycle during the radiation time-step, matter-, and radiation-update
steps makes the computation numerically entensive.
For a given set of input parameters,
we solve our problem on a $200\times 200$ grid, and compare the results with those obtained on low resolution, $100\times 100$ grid.
On both resolutions the solution converges towards a quasi-stationary one.
}

The BC for the matter provide $\rho$, $e$ and ${\bf v}$ as follows: At the equatorial plane 
the inflow BC are adopted, having always $v_{z, \rm in}>0$. An outflow, i.e. $v_{z}<0$ through the equatorial plane is not permitted. Notice that we allow for $v_{z, \rm in}$ to be arbitrary small, being numerically limited to a fraction of $v_{\rm s,g}$,
the sonic velocity calculated using only $P_{\rm g}$.
During numerical calculations the value of $v_{z, \rm in}$ mirrors the value of the velocity found in the adjacent cell inside the computational domain. 
Thus,  $v_{z, \rm in}$ adjusts in the process of computation. 
The difference of our approach from
others often adopted in simulations of accretion disk winds is that we obtain $v_{z, \rm in}$ and thus the mass-loss rate, $\dot{M}$ self-consistently. 
A power-law distribution for the density is assumed at the equator:

\begin{equation}\label{DenLaw}
\rho(0, x) = x^{-d}\mbox{,} 
\end{equation}
At all other boundaries including the inner, left boundary, outflow BC are adopted.
The azimuthal component of the velocity $v_{\phi}$ is assumed to be Keplerian at the equatorial plane.

For the radiation at the left 
boundary which is located at the distance $R_{0}$ from the BH, we specify the distribution of energy density

\begin{equation}\label{ELaw}
E(z,x_{0})  = E_{x0}  \,z^{-\epsilon} \mbox{,} \label{Elin}
\end{equation}
where $E_{x_{0}} = E(z=0, R_{0})/E_{0}$.
At the equatorial plane, the disk ''photospheric'' conditions are mimicked using the effective temperature, $T_{\rm eff}$, which provides 

\begin{equation}\label{bc-equator}
F_{z}(z_{0}, R) = -D \,dE/dz = \sigma T^{4}_{\rm eff}\mbox{,}
\end{equation}
where $T_{\rm eff}$ is calculated from
a  self-consistent ''photospheric'' boundary condition, i.e.
$T_{\rm eff} = T(z_{0}, R)$.
The free-streaming boundary conditions are assumed at 
the upper boundary, located at $z_{1}$ and at the right boundary at $R_{1}$.

Thus,
equations (\ref{eq21})-(\ref{eq24}) with boundary conditions (\ref{DenLaw})-(\ref{bc-equator}) are integrated in time until a quasi-static solution is found.

\section{Results}\label{results}

One of the primary objectives of this paper is to prove by multidimensional radiation hydrodynamics calculations the suggestion made in Paper I,
that the structure which is usually referred to as an AGN torus is better described in terms of a radiationally supported flow rather than as a quasi-static obscuring torus. 

To approach this goal 
a set of models is calculated from the first principles of radiation hydrodynamics. 
The structure we model is represented by an extended, rotating ring of radiatively-dominated plasma
in the $\{z,R\}$ plane in cylindrical coordinates.
Our calculations are performed in 2.5D meaning that we are able to calculate arbitrary 2D, time-dependent distributions of gas and radiation
and to 
model an arbitrary axially-symmetric (i.e. $\partial_{\phi}\equiv 0$) velocity field.
Apart from the fundamental properties of the torus such as the distribution of mass, radiation energy density and velocity, 
the mass-loss rate and the torus mass are readily obtained from such calculations.

\subsubsection{Basic parameters}
The BH mass, $M_{\rm BH} =1\times 10^{7}\ M_{\odot}$, and the size of the obscuration region $R_{0}=1$pc are fixed at these values in all our models. Also
the models are characterized by the Thomson optical depth of the torus, which is calculated at the inclination $\theta=\pi/2$ from the z-axis: $\tau_{\rm T}=\int_{0}^{\infty}\, \kappa_{\rm T}\rho \,dR$, and by 
parameter $\Gamma$ which measures the intensity of the illumination of the inner face of the torus by UV and soft X-rays.
The effective temperature of the conversion layer, $T_{\rm eff}$ is found then from (\ref{Teff00}), and we also adopt $T_{0} = T_{\rm eff}$ in (\ref{param1}).
The parameter $d=0.5$ enters the equatorial distribution of density (\ref{DenLaw}), and parameters $E_{x0}=1$, and $\epsilon=0.1$ shape the distribution of $E$ in (\ref{Elin}) and are fixed in all models.

The dust opacity $\kappa$ is fixed at a constant value throughout the computations. The latter approximation may create an artificial situation when a 
very low density wind with $n\ll n_{0}$ is accelerated to very high velocities.
This wind is an artifact of the adopted approximation: it has a negligible column density and a negligible mass-flux, and in a more detailed calculation, when $\kappa$ is allowed to depend on density such wind would not exist. 
It is convenient to introduce the following definitions: the average bulk velocity of the flow $\displaystyle \langle v \rangle = {\int_{V}\,\rho v dV} / {\int_{V}\, \rho dV}$, where $V$ is the total volume occupied by the flow; and the maximum velocity of the dense wind: $v^{*}_{\rm max}$ found throughout the flow given the condition
$\rho(z,R)>\rho_{\rm th}$ is satisfied. 
Experimenting with various threshold values $\rho_{\rm th}$ we found that if $\rho_{\rm th}$ is in the range $\rho_{\rm th}=0.01-0.001\rho_{0}$ the value of $v^{*}_{\rm max}$
remains almost unchanged. 
Notice that
such a definition of $v^{*}_{\rm max}$ gives results which are approximately in accord with estimates based on the kinetic output of the wind $v_{\rm kin} \simeq 2 L_{\rm kin}/{\dot M}$, where $\displaystyle L_{\rm kin}=\int_{\Sigma} \rho v^{3}/2\, d\Sigma$ is the kinetic luminosity of the wind, and $\Sigma$ is the outer boundary of the computational domain.

\begin{tabular}{c c c c c c c c c c c}
Model & $\Gamma$ & $R_{0}$ & $\tau_{\rm T}$ & $n_{0}$ & $\langle v \rangle$ &$v^{*}_{\rm max}$  & 
$L_{\rm kin}  $ & $L_{\rm bol} $ & ${\dot M} $\\
\hline
\hline
$1$ & 0.5 & 1 & 0.53& $3\cdot 10^{5}$ & 138.4 & $ 666.61 $ & $ 1.23\cdot 10^{41} $ & $6.24\cdot 10^{44}$ & $1.71$\\
$2$ & 0.3 & 1 & 0.53& $3\cdot 10^{5}$ & 104.87&  560.27 &  $ 6.38 \cdot 10^{40} $ & $3.74\cdot 10^{44}$ & 1.38\\
$3$ & 0.1 & 1 & 0.53& $3\cdot 10^{5}$ & 55.82&  373.46 &  $ 1.48 \cdot 10^{40} $ & $1.24\cdot 10^{44}$ & 1.85\\
$4$ & 0.05 & 1 & 0.53& $3\cdot 10^{5}$ & 36.63 &  282.23 &  $ 6 \cdot 10^{39} $ & $6.25\cdot 10^{43}$ & 0.64\\
$5$ & 0.8 & 1 & 1.8& $1\cdot 10^{6}$ & 112.62 & $ 765 $ & $ 4.11\cdot 10^{41} $ & $9.99\cdot 10^{44}$ & $5$\\
$6$ & 0.5 & 1 & 1.8& $1\cdot 10^{6}$ & 90.55& $ 513.42$ &  $ 2.56 \cdot 10^{41} $ & $6.24\cdot 10^{44}$ & 4.18\\
$7$ & 0.3 & 1 & 1.8& $1\cdot 10^{6}$ & 73.18&  348.45 &  $ 1.37 \cdot 10^{41} $ & $3.74\cdot 10^{44}$ & 3.37\\
$8$ & 0.1 & 1 & 1.8& $1\cdot 10^{6}$ & 59.73&  159.74 &  $ 3.09 \cdot 10^{40} $ & $1.24\cdot 10^{44}$ & 1.88\\
$9$ & 0.05 &1& 1.8& $1\cdot 10^{6}$ &42.8&  129.39 &  $ 1.31 \cdot 10^{40} $ & $6.25\cdot 10^{43}$ & 1.39\\

\hline
\end{tabular}

\tablename{ 1. Models characterized by initial parameters: $\Gamma$, $R_{0}(\rm pc)$, $\tau_{\rm T}$, characteristic density $n_{0}{(\rm cm^{-3})}$, 
and the resulting averaged flow velocity, $\langle v \rangle(\rm km\,s^{-1})$, the averaged maximum velocity, $v^{*}_{\rm max} (\rm km\,s^{-1})$,
the kinetic and bolometric luminosities, $L_{\rm kin}{\rm (erg\,s^{-1}) }$, $L_{\rm bol}{\rm (erg\,s^{-1}) }$, and mass-loss rates 
${\dot M} (M_{\rm \odot}\,{\rm yr^{-1}} )$.
}

\begin{figure}[htp]
\includegraphics[width=450pt]{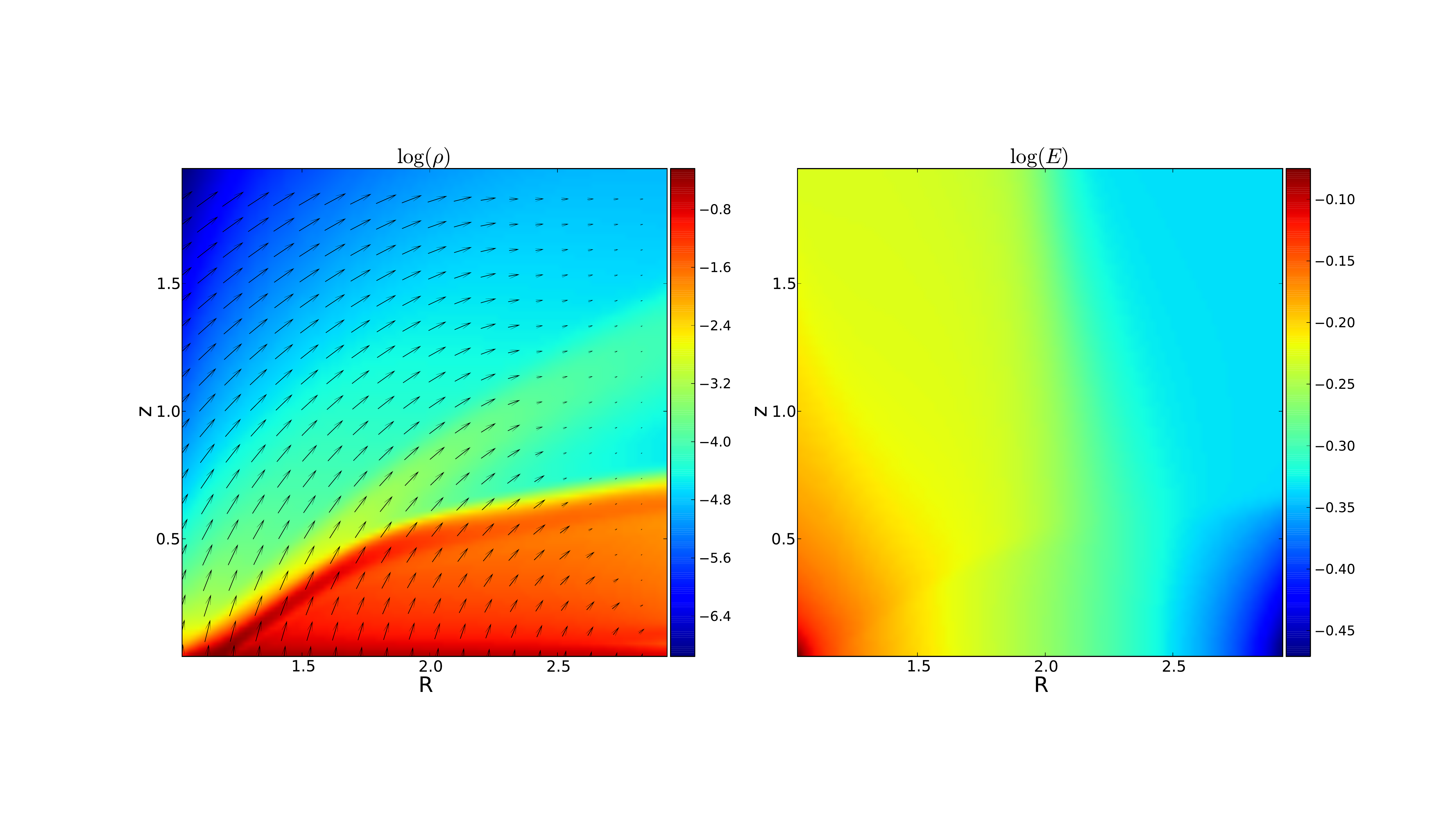}
\caption{Model 1. Color-intensity plots of the dimensionless density, $\rho$ (left) and dimensionless infrared radiation energy density, $E$ (right),
and the superimposed velocity field. Axes: distance in parsecs.
}\label{fig1}
\end{figure}

\begin{figure}[htp]
\includegraphics[width=420pt]{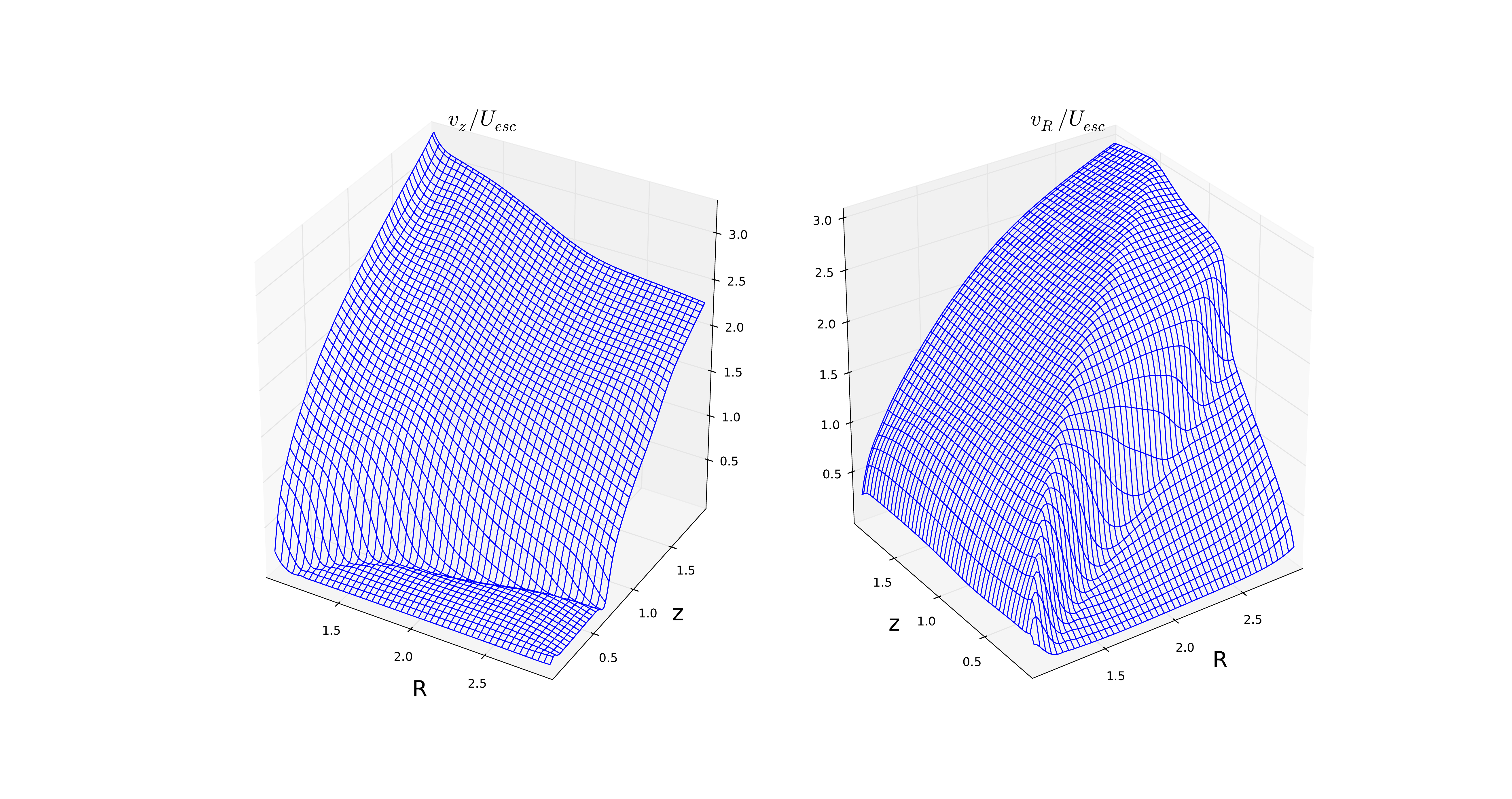}
\caption{Model 3: the surface plot of the $z-$ and $R-$ velocity components, where $U_{\rm esc}$
is the escape velocity. 
Horizontal: $R$: distance from the BH in parsecs;
$z$: distance from the equatorial plane in parsecs;
}\label{fig2}
\end{figure}

\begin{figure}[htp]
\includegraphics[width=420pt]{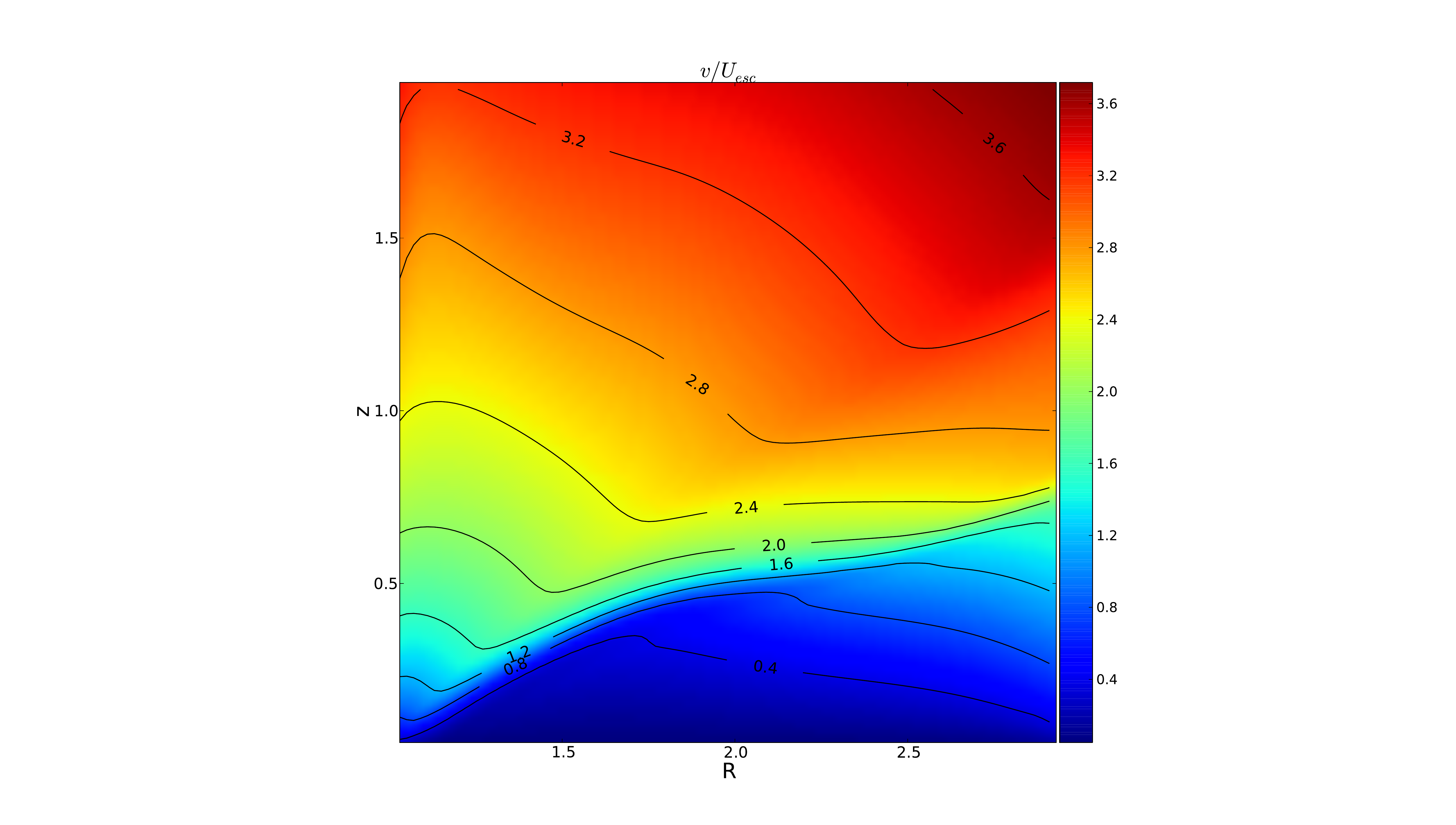}
\caption{
Model 3:  Color-intensity plot and contours of the total velocity $v =( u_{z}^{2} + u_{R}^{2}) )^{1/2}$,
where $U_{\rm esc}$
is the escape velocity.  Horizontal: $R$: distance from the BH in parsecs;
$z$: distance from the equatorial plane in parsecs;
}\label{fig21}
\end{figure}

Table 1 describes the set of calculated models, summarizes the governing parameters and outlines the most important results. 
The models are divided into two broad categories with respect to the total Thomson optical depth at the equator: marginally 
optically thin $\tau_{\rm T}\simeq 0.53$ and optically thick $\tau_{\rm T}\simeq 0.8$. For simplicity, we refer to the first category as type I models and 
to the second as type II.  After $t \simeq 3\,t_{0}$ dynamical times, where $t_{0}\simeq 1.5\times 10^{11}\, r_{\rm pc}^{3/2} M_{7}^{-1/2}$s, the quasi-equilibrium solutions are found for all models. 
{\mybf
Characteristic distributions of $\rho$, $E$ and ${\bf v}$ for type I models are shown in Figure \ref{fig1}-\ref{fig21}, and for type II models in Figure \ref{fig3}-\ref{fig51}.
}

\subsubsection{The kinetic energy of the wind}

The first apparent trend with respect to the energy budget is that pumping radiation energy into the flow (larger $\Gamma$) results in a larger kinetic 
output, $L_{\rm kin}$ in such models. 
For example we have $L_{\rm kin}/L_{\rm bol}=9.6\cdot 10^{-5}$ for Model 4, with $\Gamma=0.05$ 
while for Model 1 with $\Gamma=0.5$ we get $L_{\rm kin}/L_{\rm bol}=2\cdot 10^{-4}$. 

Similar comparison for marginally optically thin models with similar $\Gamma$ gives $L_{\rm kin}/L_{\rm bol}=2\cdot 10^{-4}$ for Model 9,  and $L_{\rm kin}/L_{\rm bol}=4.1\cdot 10^{-4}$ for Model 6.
A note of caution: the mass-loss rate, $\dot{M}$ in Table 1 reflects only the formal integral over the mass-flux at the boundary of the computational domain, which may not reflect the true escape of the gas to infinity. To illustrate this, in the following we show that most of the gas remains gravitationally bound to the BH.

\begin{figure}[htp]
\includegraphics[width=450pt]{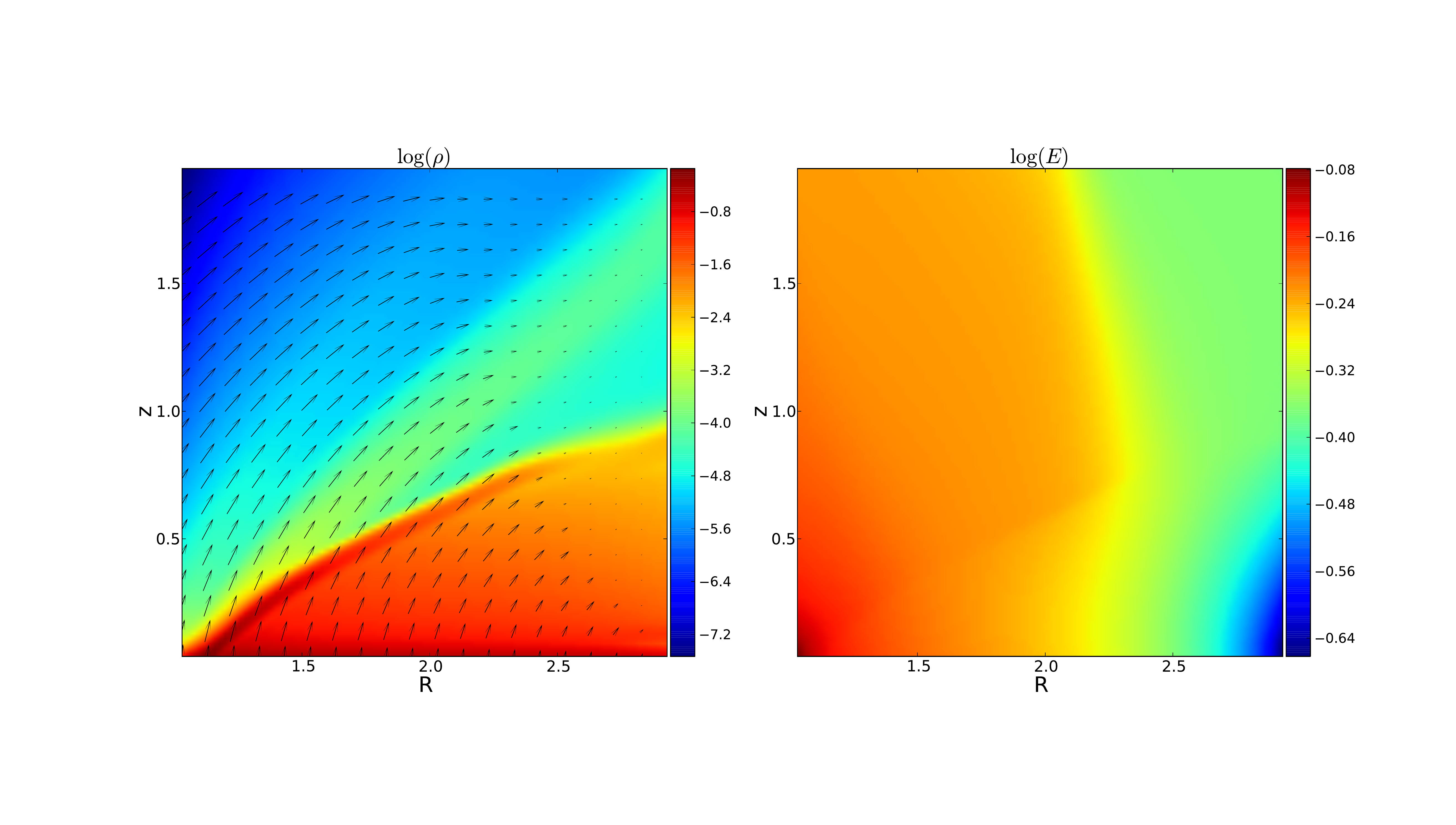}
\caption{Model 6. Color-intensity plots of the dimensionless density, $\rho$ (left) and dimensionless infrared radiation energy density, $E$ (right),
and the superimposed velocity field. Axes: distance in parsecs.
}\label{fig3}
\end{figure}

\begin{figure}[htp]
\includegraphics[width=450pt]{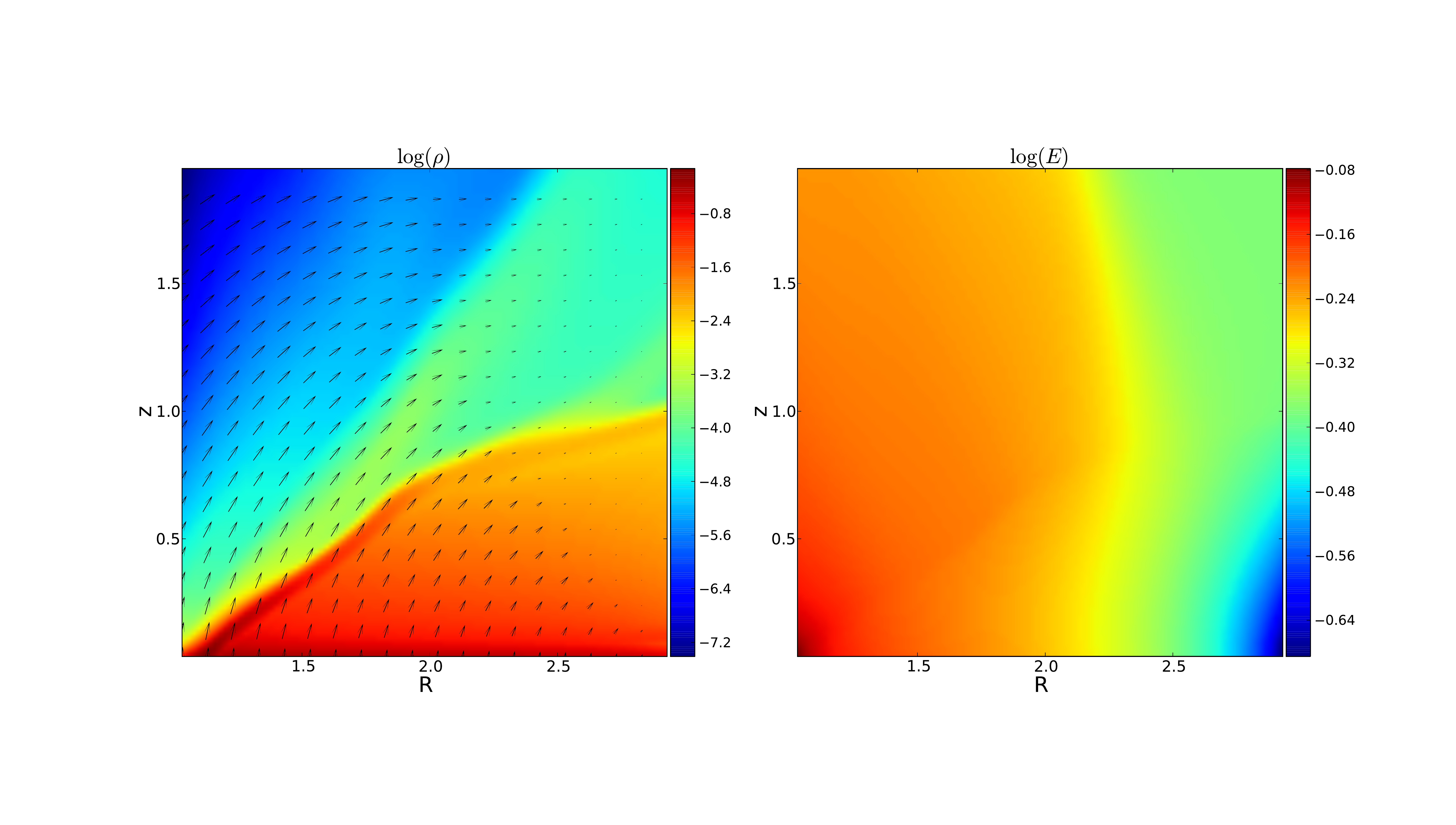}
\caption{Model 7. Color-intensity plots of the dimensionless density, $\rho$ (left) and dimensionless infrared radiation energy density, $E$ (right),
and the superimposed velocity field. Axes: distance in parsecs.
}\label{fig4}
\end{figure}

\begin{figure}[htp]
\includegraphics[width=420pt]{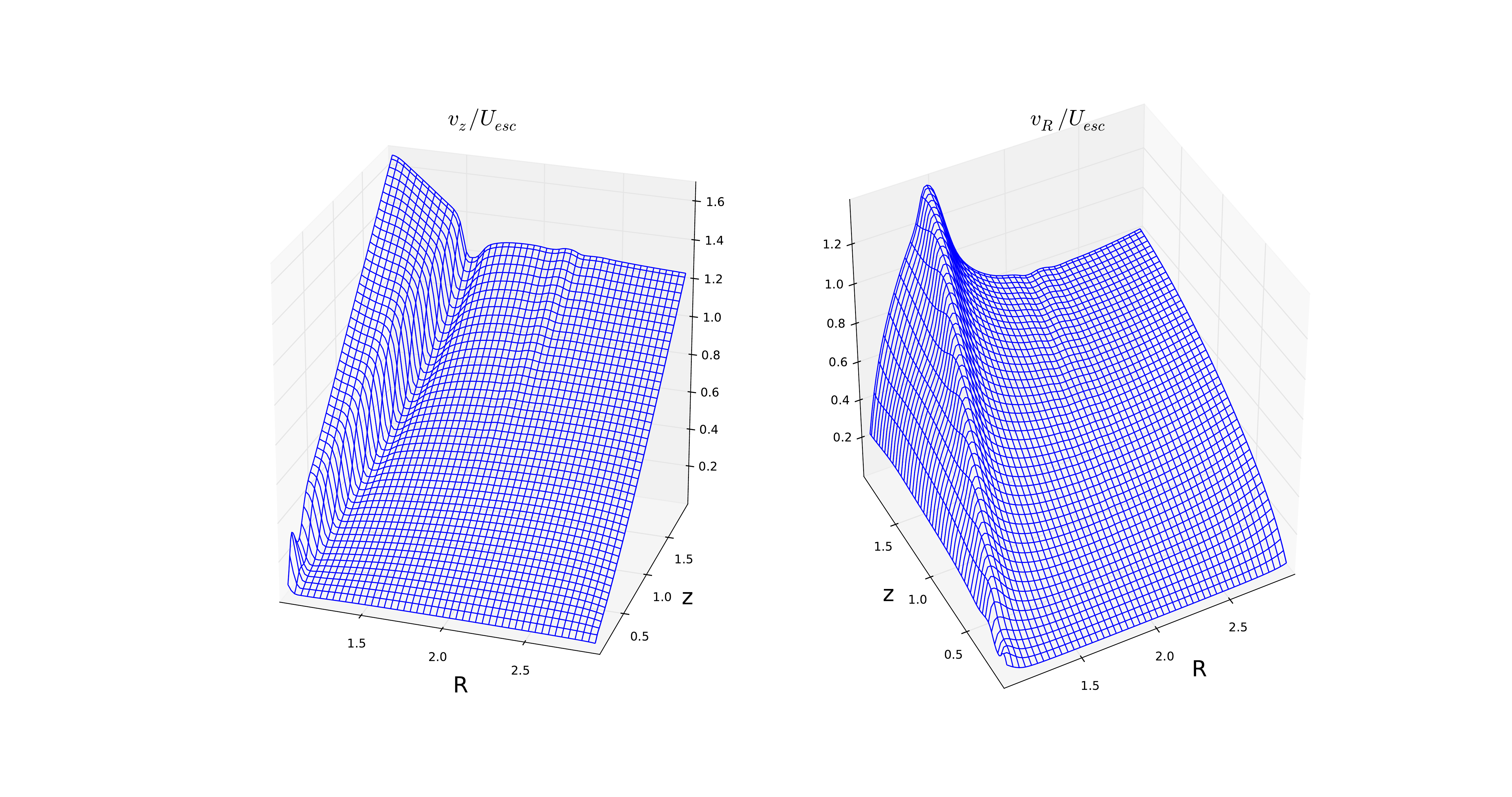}
\caption{Model 9: the surface plot of the $z-$ and $R-$ velocity components, where $U_{\rm esc}$
is the escape velocity. 
Horizontal: $R$: distance from the BH in parsecs;
$z$: distance from the equatorial plane in parsecs;
}\label{fig5}
\end{figure}

\begin{figure}[htp]
\includegraphics[width=420pt]{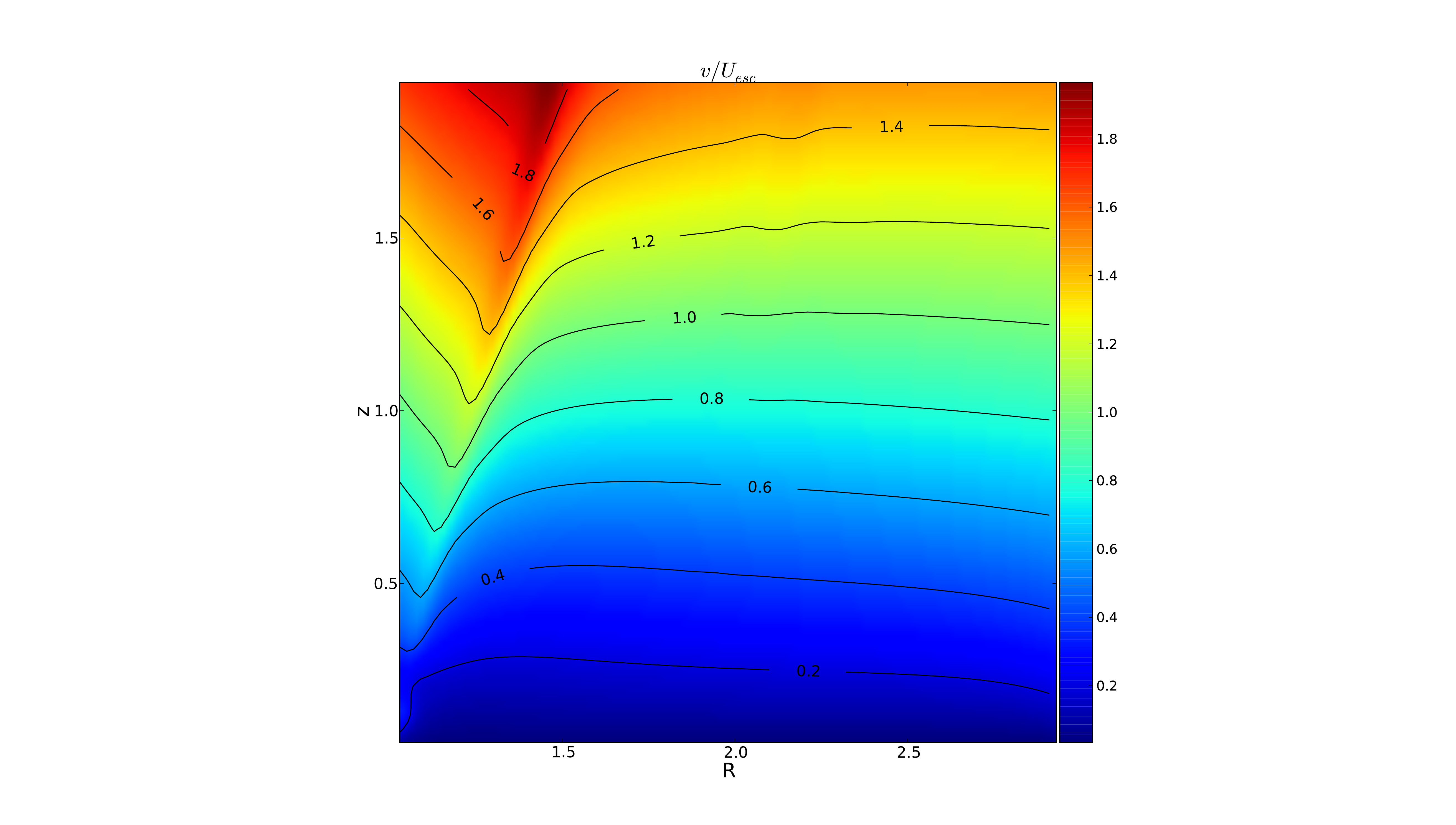}
\caption{
Model 9:  Color-intensity plot and contours of the total velocity $v$,
where $U_{\rm esc}$
is the escape velocity.  Horizontal: $R$: distance from the BH in parsecs;
$z$: distance from the equatorial plane in parsecs;
}\label{fig51}
\end{figure}

Figure 1 shows the color-intensity plot of $\rho$ (with superimposed velocity field ) and $E$ from Model 1. The effective temperature of the radiation
decreases from its maximum at the left boundary relatively smoothly, while the 
distribution of $\rho$ demonstrates a pronounced 
disk-like structure. 

This is one of the most important results of the current work: the self-consistent distribution of density naturally evolves into a geometrically thick disk-like structure.
Notice that only the density BC at the equator and the radiation energy at one boundary are provided.
The aspect ratio of the resultant toroidal structure is $h/R\sim1$, where $h(R)$ is the vertical extent of the disk.
Formation of a  geometrically thick disk is observed in other models which have higher $\tau_{\rm T}$: 
in Figure \ref{fig3} for Model 6, and in Figure \ref{fig4} for Model 7.

\subsubsection{The structure of the velocity field in the torus}
The analysis of the 
velocity field shown by arrows in Figures 1,3, and 4 supports the basic hypothesis made at the beginning of this paper: the torus is formed by gas which is not in a static equilibrium. It is not surprising that the characteristic density, $n_{0}$ (or alternatively $\tau_{\rm T}$) is one of the most important driving parameters, which critically influences the
distribution of the radiation energy density and to a large extent determines the topology and magnitude of the velocity. 
{\mybf
The most important observation is that the disk-like structure seen in density plots represents a portion of a wind which is mostly subsonic. This is a striking analogy with 
models of outflowing stellar atmospheres driven by radiation pressure in the continuum \citep{BK-Dor99} where the inner subsonic part of the wind is usually the most dense one and the transition to $\tau \lesssim 1$ happens soon after the sonic point is reached.

Figures \ref{fig21}, and \ref{fig51} show the 
color-intensity plot and contours of the total velocity $v =( u_{z}^{2} + u_{R}^{2}) )^{1/2}$. One can see that in the first case 
the geometrically thick obscuring flow does not have enough speed to escape the potential well of the BH. The fast component is formed only when density drops. Low luminosity model shown in Figure \ref{fig51} has a weak outflow without such a clear separation into low velocity and dense wind (i.e. ''atmosphere'' in a stellar analogy) and the fast wind.
}

The difference between the lower density models ($\tau_{\rm T} = 0.53$, i.e. Figure \ref{fig1}) and higher density ones ($\tau_{\rm T} = 1.8$, i.e. Figure \ref{fig4}) is that in the latter there is a larger portion of the wind where the velocity is comparable to the escape velocity of the gas, 
$U_{\rm esc}= (GM_{\rm BH}/R_{0})^{1/2}\simeq 207 \, M_{7}^{1/2} R_{\rm pc}^{-1/2}{\rm km\, s^{-1}}$.
Most of the torus mass is participating in the intensive global motion, though most of the gas
remains in the potential well of the BH.
Notice that
we do not see a  quasi-static disk in our solution at any stage of our simulation. However if $\Gamma$ is not too small 
the velocities in a denser, disk-like part of the flow are such that $v_{R}\ll v_{z}\sim U_{\rm esc}$  (c.f. Figure \ref{fig2}). A similar situation is observed in the most of our models except for type II models with $\Gamma=0.05$. In the latter case (c.f. Figure \ref{fig5}), $v_{R}\sim v_{z}$ everywhere in the computational domain, except for the region close to the conversion layer,
where  $v_{z} > v_{R}$.
The results show that the
average velocity of the obscuring flow, $\langle v \rangle$ is almost always comparable but smaller than the
escape velocity: for example, $\langle v \rangle = 0.53\, U_{\rm esc}$ for Model 5; $\langle v \rangle=0.43\, U_{\rm esc}$ for the Model 6; and
$\langle v \rangle = 0.28\, U_{\rm esc}$ for Model 8.

For models with $\tau_{\rm T} = 0.53$
the maximum velocity reaches $v^{*}_{\rm max} = 3.17\, U_{\rm esc}$ for Model 1, $v^{*}_{\rm max} = 2.66\, U_{\rm esc}$ for Model 2, dropping further until it is $v^{*}_{\rm max} = 1.34\, U_{\rm esc}$ for Model 4. 
Comparable results are obtained for models with $\tau_{\rm T} = 1.8$: the maximum $v^{*}_{\rm max} = 3.64\, U_{\rm esc}$ is obtained for Model 5.
A $\Gamma$ is reduced so does the $v^{*}_{\rm max}$, being $v^{*}_{\rm max} = 2.44\, U_{\rm esc}$ for Model 6, and finally 
$v^{*}_{\rm max} = 0.76\, U_{\rm esc}$ is obtained for Model 8.

\subsubsection{The column density and optical depth}
Comparing various models at $\theta=45^{\circ}$ one can see that  
Models 1-4~($\tau_{\rm T} = 0.53$) do not have enough column density to provide 
Compton-thick obscuration at this inclination.
On the other hand, at higher inclinations both sets of models provide a noticeable extinction of the light
from the supermassive BH. Notice that we do not account for 
the possible nonzero optical depth provided at smaller radii by any sort of additional wind, such as MHD 
wind or warm absorber flow. Such flows can add to column densities enough to provide a wind optical depth, $\tau_{\rm w}\lesssim 1.$

The column density in Model 1 increases from $N_{\rm col} = 9.6\cdot 10^{18}\, {\rm cm}^{-2}$, at the inclination, $\theta=45^{\circ}$ to
$N_{\rm col} = 1.49\cdot 10^{20}\, {\rm cm}^{-2}$ at $\theta=65^{\circ}$, and $N_{\rm col} = 6.8\cdot 10^{23}\, {\rm cm}^{-2}$ at $\theta=90^{\circ}$.
The column density in Model 3 increases from $N_{\rm col} = 8\cdot 10^{18}\, {\rm cm}^{-2}$ at $\theta=45^{\circ}$ to
$N_{\rm col} = 1.3\cdot 10^{20}\, {\rm cm}^{-2}$ at $\theta=65^{\circ}$, and $N_{\rm col} = 7.4\cdot 10^{23}\, {\rm cm}^{-2}$ at $\theta=90^{\circ}$.
The column densities for the denser Models 5-9 ($\tau_{\rm T} = 1.8$) are higher. 
For example, in Model 5 one obtains:
$N_{\rm col} = 8.8\cdot 10^{18}\, {\rm cm}^{-2}$ at $\theta=45^{\circ}$; $N_{\rm col} = 3\cdot 10^{20}\, {\rm cm}^{-2}$ at $\theta=65^{\circ}$;
and $N_{\rm col} = 2.3\cdot 10^{24}\, {\rm cm}^{-2}$ at $\theta=90^{\circ}$.

As the luminosity is reduced by almost an order of magnitude, such as in Model 8, part of the wind closer to the disk (higher inclinations) becomes
denser, and the column density increases correspondingly, to
$N_{\rm col} = 8.2\cdot 10^{21}\, {\rm cm}^{-2}$ at $\theta=45^{\circ}$; $N_{\rm col} = 3.8\cdot 10^{22}\, {\rm cm}^{-2}$ at $\theta=65^{\circ}$;
and $N_{\rm col} = 2.4\cdot 10^{24}\, {\rm cm}^{-2}$ at $\theta=90^{\circ}$.

In order to provide 	obscuration, for example, at $30^{\circ}$ away from a disk plane
the AGN torus should have equatorial densities at least of the order of $10^{6} {\rm cm^{-3}}$

{\mybf We also calculated the infrared (i.e. with respect to dust) optical depth, $\tau_{{\rm IR},z}$ in z-direction as measured from the 
upper boundary of the domain. 
Models 5-9 undergo the transition from optically thick to marginally optically thin ones at $z \simeq 0.25-0.5$ with a shape which closely follows density distribution. The inner low density funnel is always optically thin.
Above this region there is an extended marginally optically thick region. Lower density models 1-4 have optically thick but geometrically thin 
disk, of of vertical extent $z \lesssim 0.2$. The rest of the wind has $\tau_{{\rm IR},z} \lesssim 1$ closely tracing the distribution of $\rho$. 

The distribution of $v_R$ (e.g. Figure \ref{fig2} ), demonstrates that  $v_R$ is increasing in a quasi-monotonic way along the 
spherical radius, $r$. This is  analogous to a one-dimensional, radiation-driven stellar wind. On the other hand, the acceleration region 
is clearly seen in the plots of $v_z$. At a given height, $z$ the region of a rapid increase of $v_{z}$ extends in radial direction until approximately 
the optical depth in the infrared (as measured from the {\it right} boundary), $\tau_{{\rm IR},R} \gtrsim 1$. When $\tau_{{\rm IR},R}$ drops, the radiation 
flux becomes free streaming and no significant lifting force in $z$ -direction is generated.

} 

\subsubsection{The mass of the torus}
As the characteristic density, $n_{0}$ increases by a more than an order of magnitude when we go from Models 1-4 to
Models 5-9,
the mass of the torus, $M_{\rm tor}$ increases only by a factor of 3: from $1.3\, \times 10^{4}\,M_{\odot}$ (Models 1-4),
to $4.5\, \times 10^{4}\,M_{\odot}$ (Models 5-9). 
These numbers provide probably the first self-consistent estimates of the torus mass.
From the above one can see that, though $M_{\rm tor}\ll M_{\rm BH}$, the self-gravity may be important inside the torus body closer to the equatorial plane where densities are higher.  
{\mybf We performed test calculations and found that basic parameters of the torus are almost unaffected by the choice of the density distribution parameter, $d$ in (\ref{DenLaw}). The more important driving parameter which sets the scaling for mass-loss rate, velocity field, and the torus mass is the characteristic density $n_{0}$
(or $\tau_{\rm T}$).
}

\subsubsection{Relevance to previous work and limitations of the present model}
It is instructive to contrast 
the approach of the current work to that of Paper I.
Compared to the current work, Paper I contained
several simplifying assumptions, 
from which the most stringent were the following two:  a) A monotonically accelerating wind with $v_{z}>0$ was assumed to exist everywhere in the 
domain and consequently no inflow solutions were permitted, and b) only the $z$-component of the velocity was taken into account.
In Paper I
the 2D distribution of $E$ was calculated from the  condition $\nabla\cdot {\bf F}=0$ assuming $t_{\rm r} \ll t_{\rm f}$ (see the discussion 
in Section \ref{condtorus}).
Additionally, in Paper I the equations included only the momentum and continuity equation for the matter, and no gas pressure nor the energy equation for the matter were considered. Essentially, in the previous studies the radiation force was calculated from a solution of a 2D diffusion problem assuming radiative equilibrium, and only after that plugging a $z$-component of this force into a 1D wind problem in the $z$-direction.

In this paper we solve the full system of radiation-hydrodynamics equations, adopting from Paper I only the description of the boundary conditions.
In the present work $v_{\rm max}$ is systematically larger than $v_{\rm max}= v_{z, \rm max}$ from Paper I. 
In Paper I the streamlines were directed along the $z$-axis, but in 2D they can bend, return or be tightly packed towards the equatorial plane. Thus the 
addition of another degree of freedom in the present work results
in smaller mass-loss rates.

For example, comparing models with $\Gamma=0.8$, $\tau_{\rm T}\simeq 2$,  we see
that $v_{\rm max}$ is considerably larger in full RHD modeling:  $v_{\rm max}^{*}\simeq 160{\, \rm km\,s^{-1}}$ versus $v_{\rm max}\simeq 760{\, \rm km\,s^{-1}}$
where results from Paper I are marked by a ''$^{*}$''. This illustrates our previous statements about the very low density and high velocity portion of the wind.
The velocities in the denser parts are in accord with previous results: 
$\langle v \rangle \simeq 112{\, \rm km\,s^{-1}}$. Removing the restrictions of Paper I, the most stringent being the $z$-only motion of the gas, also results in higher kinetic energy of the wind: $L_{\rm kin}\simeq 5\times 10^{41} {\, \rm erg\, s^{-1} }$ versus 
$L_{\rm kin}^{* }\simeq 5.5 \times 10^{39} {\, \rm erg\, s^{-1} }$. In both sets of models $L_{\rm kin}$ is still a tiny fraction of $L_{\rm bol}$.
The mass-loss rate in the full case is lower than in Paper I: for the same set of parameters as above ${\dot M}$ is reduced from
$9.5\, M_{\odot}\,{\rm yr}^{-1}$ to $5\, M_{\odot}\,{\rm yr}^{-1}$. Lower density models have a similar trend of increasing the $v_{\rm max}$ and 
$L_{\rm kin}$ compared to the previous studies and reducing ${\dot M}$. 
For the models with $\Gamma=0.3$, $\tau_{\rm T}\simeq 0.5$ the latter is again reduced by almost 50\%, from
$2.76\, M_{\odot}\,{\rm yr}^{-1}$ to $1.38\, M_{\odot}\,{\rm yr}^{-1}$.

{\mybf
One of the serious limitations of our model is due to the non-vanishing dust opacity. In more realistic simulations the very low density parts of the wind, i.e.
the funnel seen in the $\rho$ plots will probably not exist. The dust there will most likely sublime and the funnel will be filled with much hotter gas. 
One can expect a picture will resemble an X-ray evaporative flow of \cite{Dorodnitsyn08b} but instead of a quasi-static torus there will be an X-ray induced  evaporation of a dense infrared driven flow. Even without X-rays in the very low density part the dust will decouple from the gas, breaking the one-fluid hydrodynamics approximation, and thus such dust even if survived will be quickly blown away.
Summing up we believe that the high velocity and low density component is most likely an artifact of our simplifying assumptions.
We will address this in future work.

The finite optical depth provided by some other gas/wind at smaller radii is also not taken into account. For example, if a broad absorption line (BAL) wind is formed closer to BH and the transverse (approximately) optical depth, $\tau_{\rm tr}$ between the corona and the dusty, infrared-dominated outflow is large, then nothing will be left for the torus. 
On the other hand to have $\tau_{\rm tr}>1.$ requires a wind which is more massive than we typically observe in BALs. 
In any case if $\tau_{\rm tr} \gg 1.$ the model developed in this paper is not applicable.
In case of no external heating, the only source of radiation is the accretion disk itself. This should be addressed in a future work. Help may come from hard X-rays which can penetrate much deeper in the torus body providing distributed sources of heat. That was shown by \cite{Chang_etal07} to be quite effective in puffing up the initially geometrically thin accretion disk. In more complete simulations, when heating by hard X-ray is implemented we expect 
more efficient acceleration at larger R.  At present our calculation gives the most conservative estimate of the 
radiation driving at large distances from the  BH. 
The problem of AGN winds is 
very non-linear and interdependent one, and 
to answer these questions global multi-group radiation-hydrodynamics simulations of AGN accretion disk + winds are required. 
}

\section{Observational consequences and conclusions}\label{conclusion}

It is widely accepted that AGN unification schemes require an obscuring toroidal structure as a basic premise.
Obscuration can be quasi-static or dynamical (winds). 
An example of the first kind of obscuration is the 
polytropic torus, which is a rotationally supported torus with the equation of state $P_{\rm g}\sim \rho^{1+1/n}$, where $n$ is the index
of the polytrope. Such a torus has been shown to be unstable to 3D non-axisymmetric perturbations by \cite{PapaloizouPringle84}, but even without such a difficulty it is likely inappropriate as an AGN torus prototype. The temperature in a gas-pressure-supported torus is of the order of the virial temperature which is 
$2.6 \times 10^{6}\,M_{7}/r_{\rm pc}$ K. That is far too high to be reconciled with the existence of dust which requires temperatures of the order of $100-1000$ K.
Obscuration can be clumpy/cloudy but we believe that this is the level of complexity which is of the next order compared to the 
fundamental question approached in the current paper. 

The problem of AGN unification via toroidal obscuration can be formulated in the following form: what supports the torus against vertical collapse to a geometrically thin state and thus maintains its aspect ratio $h/R\sim 1$? 
Infrared pressure on dust grains seems to be the best candidate.  
In Paper I it was shown that if the temperature inside the torus is of the order of
$T_{\rm vir, r}   \simeq 312\,(n_{5} M_7 / {r_{\rm pc} })^{1/4}-987\,(n_{7} M_{7}/ r_{\rm pc})^{1/4} \,{\rm K}$
an equilibrium between radiation pressure, rotational and gravitational forces cannot be maintained, which results in formation of an outflow resembling that of an 
accretion disk wind.

The conversion of external UV and X-ray radiation into IR pumps the torus with IR photons. Internal temperatures of  a ${\rm few}\times10^{2}$
K provide conditions for the extensive presence of dust which results in a strong coupling between the gas and IR photon field due to the high opacity of dust to IR radiation.
In the bulk of the infrared supported torus
the infrared radiation pressure $\Pi \gg P_{\rm g}$, with the possible exception of the equatorial region.
Correspondingly, the torus problem must be formulated in terms of equations of radiation hydrodynamics. In the current paper we solved a full system of such equations and found that if external BH luminosity exceeds $\sim 0.05\,L_{\rm edd}$
an outflow is created, and that significant masses are involved in global motions. 

The obtained mass-loss rates 
depend on $\Gamma=L/L_{\rm edd}$ and on the characteristic density, $n_{0}$ which scales the distribution of $\rho$ in the equatorial plane.
Generally, models with higher $\Gamma$ and with larger $n_{0}$ (or alternatively $\tau_{\rm T}$) tend to have higher mass-loss rates. However there exist an
overlap, when marginally optically thin but luminous models produce mass-loss rates similar
to the optically thick and less luminous ones. For example,
the model with $\tau_{\rm T}\simeq 0.5$, and $\Gamma=0.3$ has mass-loss rate, ${\dot M}=1.4\,M_{\odot}\, {\rm yr}^{-1}$, similar
to the model with $\tau_{\rm T}\simeq 2$ and $\Gamma=0.05$. 
Besides such overlap, in optically-thick models with higher $\Gamma$ much larger ${\dot M}$ is obtained than in optically thin ones. 
For example,
in one of the Thomson thick and luminous models with $\tau_{\rm T}\simeq 2$ and $\Gamma=0.8$ the mass-loss rate is ${\dot M}=5\,M_{\odot}\, {\rm yr}^{-1}$
Another important finding is that most of the flow has velocity $\langle v \rangle \lesssim U_{\rm esc}$, that is too small to escape the potential well  of the BH. Thus the AGN obscuring flow is better described in terms of a failed wind rather than viewed as a bipolar and 2.5D analogy of a stellar wind.

In our calculations a number of simplifications were made. 
The most restrictive one is the shape of the conversion layer, where UV and soft X-rays are converted into the infrared. 
We assume that such a conversion is happening at the vertical boundary of the computational domain while in 
real AGN two important complications arise:
i) the 
curvature of such layer is important and can only be determined during self-consistent global radiation hydrodynamics simulations, and ii) the corresponding ''photospheres'' are located at different depths which could be captured in RHD simulations with multi-group description of radiation.
Taking into account additional heating by hard X-rays would also contribute to the structure and dynamics of the obscuring flow 
\citep{Chang_etal07, ShiKrolik08}.
These effects must be incorporated into future global radiation hydrodynamics simulations.

Observationally,
tracing wind kinematics through the detection of maser emission is one way to ''see'' an AGN obscuring flow. 
From our results it follows that infrared supported flow naturally produces outflows with bulk velocities as large as 
$\sim {\rm few} \times 100\,{\rm km\, s^{-1}}$.
It is beyond the scope of this paper to calculate conditions and particular locations in the flow suitable for maser emission. 
However, our solutions allow us to predict that if such emission is observed at distances $0.4-1.5$ from a BH
and being offset from the corresponding Keplerian velocity of the equatorial disk by 
several hundreds ${\rm km\,s^{-1}}$, it may indicate an IR-driven outflow. Such evidence may already be present in
the VLBI observations of a broad bipolar outflow in Circinus galaxy in
$\rm H_{2}O$ maser emission \citep{Greenhill03}.

This research was supported by an appointment at the NASA
Goddard Space Flight Center, administered by CRESST/UMD
through a contract with NASA, and by grants from the NASA
Astrophysics Theory Program 10-ATP10-0171.
G.B-K. acknowledges the support of from the Russian Foundation for Basic Research
(RFBR grant 11-02-00602).

\bibliography{BibList-tor6}{}

\end{document}